# Hourglass pore effect and membrane osmotic diode behavior: model and simulations

Patrice BACCHIN[a,*]

[a]Laboratoire de Génie Chimique, Université de Toulouse, CNRS, INPT, UPS, Toulouse, France
*corresponding author: patrice.bacchin@univ-tlse3.fr

## Abstract

A membrane can be represented by an energy landscape that solutes or colloids must cross. A model accounting for the momentum and the mass balances on the membrane energy landscape establishes a new way of writing for the Darcy law. The counter pressure in the Darcy law is no longer written as the result of an osmotic pressure difference but rather as a function of colloid-membrane interactions. Physically, the colloid-membrane interactions are slowing down the colloid velocity thus inducing a relative fluid-colloid motion that in turn leads to the counter pressure. The ability of the model to describe the physics of the filtration is discussed in detail. This model is solved on a simplified energy landscape to derive analytical relationships that describe the selectivity and the counter pressure from ab-initio operating conditions. The model shows that the stiffness of the energy landscape has an impact on the process efficiency: a gradual increase in interactions (like with hourglass pore shape) can reduce the separation energetic cost. It allows the introduction of a new paradigm to increase membrane efficiency: the accumulation that is inherent to the separation must be distributed across the membrane.

## 1 Introduction

The transport of colloids across interfaces is still a scientific challenge meeting applications in many processes. Flow through semi-permeable membranes is a common process in living bodies (kidneys, membrane cells, etc.) and in industrial applications (filtration, desalting, etc.) [1]. Beyond these applications, the recent development of microfluidic experiments and the nano-scale engineering of interfaces have revived the question of the role played by colloid-surface interactions on transport across interfaces [2].

The transport of fluids across a membrane is classically determined with a modified Darcy law accounting for a counter osmotic pressure:

$$J = \frac{L_p}{\eta}(\Delta p - \sigma \Delta \Pi) \qquad (1)$$

where $J$ is the solvent flux through the membrane, $L_p$ is the membrane permeability, $\eta$ is the solvent viscosity, $\Delta p$ is the transmembrane pressure and $\Delta \Pi$ is the transmembrane osmotic pressure. This writing derives from the semi-empirical formulation of Kedem and Katchalsky [3] that considers non-equilibrium thermodynamics with the assumption of linearity between the fluxes and the driving forces. This law introduces a Staverman coefficient, $\sigma$, that is difficult to interpret and still the subject of debate. Physically,

Eq. 1 signifies that the intensity of the solvent flux is induced by a driving pressure which is the applied static pressure minus the osmotic pressure difference. It is therefore assumed that flows through the membrane are caused by the differences in chemical potentials occurring across the membrane.

Several points can be subject of discussion in this law. First, this law is integrated and does not enable the accounting of a local concentration variation close to the interface. For this reason, this relationship is often considered as a flow boundary condition in a filtration problem. Secondly, in this approach the membrane is discontinuously treated as an infinitively thin membrane separating two compartments: the membrane is therefore described with a partition coefficient that induces an unrealistic concentration jump at the interface. For these reasons, such way of writing cannot help to elucidate the "strange" transport mechanism of fluids at the nano-scale [4] (recently highlighted in microfluidic experiment or with nanotube, aquaporin, etc.) and therefore to progress by designing specific nanoscale molecule/pore interactions within artificial nano-pores in order to optimize the transport [5].

A significant progress could be made by a better understanding of the roles played by colloid-membrane interaction on these interfacial phenomena. This role has been qualitatively pointed out since the earlier works on Brownian diffusion at the interface [6–8]. In 70's, several authors [9,10] develop theoretical models to quantify the role of colloid-membrane interactions on the transport. However, these models were not really developed and used by the membrane community. In these last ten years, several authors pointed out again the important role played by solute or colloid-membrane interactions on the transport through a membrane [11–14]. Recently, a two-fluid model that introduces an energy landscape to account for colloid-membrane interactions has been proposed [15]. Such a model describes the dynamics of osmotic flows with a set of continuous equations [16]. Additionally, this type of model can be helpful to explore how the membrane energy landscape can change the transport efficiencies. The aim of this paper is to analyze the effect of the energy barrier profile on the efficiency of the separation with membrane processes.

## 2 Theoretical background and model development

A transport model was recently established from the momentum balance for the fluid and the colloid phase where the membrane is represented by an energy landscape [15]. The energy landscape [17] maps the colloid-membrane interaction energy (related to Gibbs free energy that will be expressed per unit of volume as a pressure, $\Pi_i$, referred to as interfacial pressure) for of all the spatial positions of the colloids at the vicinity or inside the membrane. This map represents the overall interactions between the colloids and the membrane interface (as for example DLVO and hydration forces [18,19]). This map can also account for the energy required to change the colloid shape towards a configuration that can make the passage through the pore possible (for example for deformable particles or for extensible or unfolding proteins) [20]. The following sections of this paper will establish the transport equations on an energy landscape (2.1), will present the formalism to describe a membrane with an energy landscape (2.2) and will derive analytical relationships to express the membrane selectivity and the counter pressure (2.3). A final section (2.4) will present a numerical application of the model and its validation by comparing the results with the classical Kedem and Katchalsky approach.

### 2.1 Transport of colloids and fluid

A two-fluid model based on an energy landscape enables the implementation of colloid/membrane interactions in the fluid and colloids momentum and mass balances [15]. The two-fluid (or mixture) model

[21,22] allows the velocities of the colloid phase, $\boldsymbol{u}_c$ and of the fluid phase to be defined. By addition, the velocity of the mixture phase, $\boldsymbol{u}_m$, can be determined (supplementary information S1). From the application of the momentum balance [16], a set of equation that defines these velocities can be written:

$$\frac{\phi}{V_p} \frac{\boldsymbol{u}_m - \boldsymbol{u}_c}{m(\phi)} - \nabla \Pi_{cc} - \phi \nabla \Pi_i = 0 \tag{2}$$

$$\frac{\eta_m}{k_p} \boldsymbol{u}_m + \nabla p + \phi \nabla \Pi_i = 0 \tag{3}$$

The momentum balance on the colloids phase (Eq. 2) establishes that the drag force acting on colloids due to the slip velocity (first term) is counterbalanced by the osmotic pressure gradient (second term) and the interfacial pressure gradient due to colloid/membrane interactions (third term). In the momentum mixture balance (Eq. 3), the pressure drop - due to friction between the mixture and the interface (first term) - is brought both by the fluid pressure gradient (second term) and the colloid-membrane interaction (third term). This mathematical writing thus accounts for the interactions between the different phases. Eq.3 signifies that the drag force acting on particles (**fluid-colloid** slip velocity) is balanced by either **colloid-colloid** interactions (through the osmotic pressure, $\Pi_{cc}(\phi)$) or by **colloid-membrane** interactions (through the interfacial pressure, $\Pi_i(x)$). Eq. 4 signifies that the drag forces on the interface (**fluid-membrane** dissipation) are induced by the destocking of either the fluid pressure (**fluid-fluid** interactions) or the **colloid-membrane** interactions.

In a more conventional form for membrane community, Eqs 2 and 3 express the mass flux, $j = \boldsymbol{u}_c \phi$, and the permeate flux, $J = \boldsymbol{u}_m$:

$$j = J\phi - V_p m(\phi) \left( \frac{d\Pi_{cc}(\phi)}{dx} + \phi \frac{d\Pi_i(x)}{dx} \right) \tag{4}$$

$$J = -\frac{k_p(x)}{\eta} \left( \frac{dp(x)}{dx} + \phi \frac{d\Pi_i(x)}{dx} \right) \tag{5}$$

The writing of mass flux (Eq. 4) is similar to the conventional one: the mass flux results from the interplay between an advective term, a diffusive term (linked to the collective diffusion accounting for Brownian diffusion and colloid-colloid interaction via the gradient in osmotic pressure, $\frac{d\Pi_{cc}(\phi)}{dx}$) and a migration term induced by colloid-membrane interactions (relative to the gradient of interfacial pressure, $\phi \frac{d\Pi_i(x)}{dx}$). The coupling of these three terms helps to describe the main mechanisms that occur during separations:

- the coupling of convective and diffusive –entropic contribution- fluxes describes the polarization concentration [23,24]
- the coupling of convective and diffusive – contribution induced by colloid-colloid interaction - fluxes describes the colloid flux paradox [25] and the existence a critical flux [26–28] for homogeneous liquid-solid transition [29,30]
- the coupling of convective flux and migration induced by colloid-membrane interactions describes the heterogeneous critical flux phenomena [31,32]
- the coupling of diffusive flux and migration induced by colloid-membrane interaction describes the Boltzmann exclusion that is due to the colloid-membrane interaction [16]

However, a strong difference with conventional approaches appears in the expression of the permeate flux (Eq. 5). The driving pressure is expressed as the combination of the applied static pressure with the interfacial pressure due to colloid-membrane interaction, $\phi \frac{d\Pi_i(x)}{dx}$. The colloid membrane interactions here play the role of a forcing term on the momentum equations of the fluid flow, similarly to the Force Coupling Method used to describe multi-phase flows [33–35]. This expression, based on a mechanical approach, presents significant differences with the classical thermodynamic approach (Eq. 1) but is close to the Kedem and Katchalsky equations solved through a potential-energy profile [20]. It should be noted that, in the limit of $j = J\phi$ or $\boldsymbol{u}_m = \boldsymbol{u}_c$ that corresponds to negligible colloid-fluid drag force, the colloid membrane interaction are equal to the gradient in osmotic pressure, $\phi \frac{d\Pi_i(x)}{dx} = -\frac{d\Pi_{cc}(\phi)}{dx}$. Under this assumption, the integration of Eq. 5 thus leads to Eq. 1. The main discrepancies occurring for others conditions will be discussed in detail throughout the paper.

To determine the membrane transport, the set of equation (Eqs. 4 and 5) must be solved in the x direction, perpendicular to the membrane surface, i.e. all along the polarization and the membrane layers. The interfacial pressure variation, $\Pi_i(x)$, defines an energy landscape that represents the colloid-membrane interactions which in turn control the membrane transport through the term $\phi \frac{d\Pi_i(x)}{dx}$. This mechanical description is close to the thermodynamical model developed from a Kedem and Katchalsky approach through a potential energy profile [9].

## 2.2 Membranes as an energy landscape

The membrane is represented by an energy map that defines the energy required to pass though the membrane. The energy landscape can represent different separation pathways (Fig. 1) with the possibility to account for patchy interactions. In landscape diversity, an infinitely thin and totally retentive membrane is represented by an energy landscape being a Dirac delta function (the limit of the normal distribution plotted in Fig. 1a). In order to account for the transport inside the membrane thickness, the membrane can be represented as a pulse of interfacial pressure (Fig. 1b). In this case, the increase in interfacial pressure on both membrane side represent the partition induced by the presence of the membrane. This model is then close to the one used by modeling approaches considering partitions at both side of an internal pore where a coupling between convection and diffusion takes place [36]. But the energy landscape can also describe less classical separation pathways, which can result from a nanoscale mosaic of attractive and repulsive interactions. New opportunities are offered by nanofluidic systems to create and investigate more complex interaction pathways that can bio-mimick biological membranes. Asymmetric exclusion can be accounted for (Fig. 1c) to describe nanofluidic osmotic diodes [37]. The interaction pathway can also exhibit internal energy wells (Fig. 1d) which can depict either double-skin membranes, or, the presence of macro-pores inside the membranes. The stiffness of the interaction pathway can also vary to represent, for example, membranes with a porosity gradient (Fig. 1e).

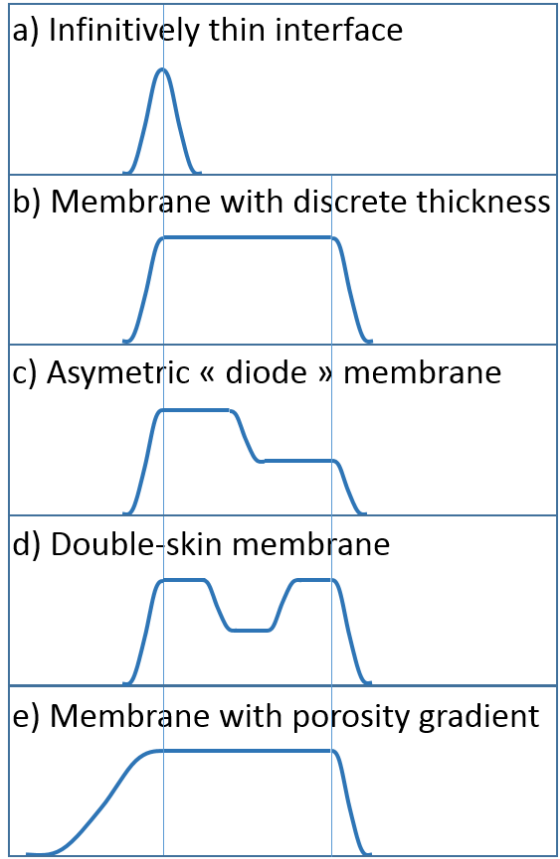

*Figure 1: Energy landscape representation of membrane diversity.*

In this paper, the energy map for the membrane will be modeled in its simplest form in order to analyze the essential ingredients of the model. The membrane is defined by two exclusion layers (where a constant interfacial gradient takes place) bordering a zone with a constant interfacial pressure (the gradient is zero and does not induce any forces in the momentum balances). This case can represent the energy landscape plotted in fig. 1b and 1e. The main parameters of the energy map (Fig. 2) are the maximum value reached in the membrane, $\Pi_{i\,max}$ (that can be expressed in a non-dimensional way, $Ex = \frac{V_p \Pi_{i\,max}}{kT}$) and the length of the ramp, $\delta_{EX}$. In the case of an infinitesimally thin membrane (demonstration in supplementary information S2), the exclusion number can be linked to a partition coefficient, K, translating the Boltzmann exclusion induced by the energy map, $K = e^{-Ex}$ [16]. The model will be applied through a series of composite layers (Fig. 2) that represent the boundary layer thickness where concentration polarization occurs, the exclusion layer at the membrane entrance, the membrane core thickness, the exclusion at the membrane outlet and the permeate bulk.

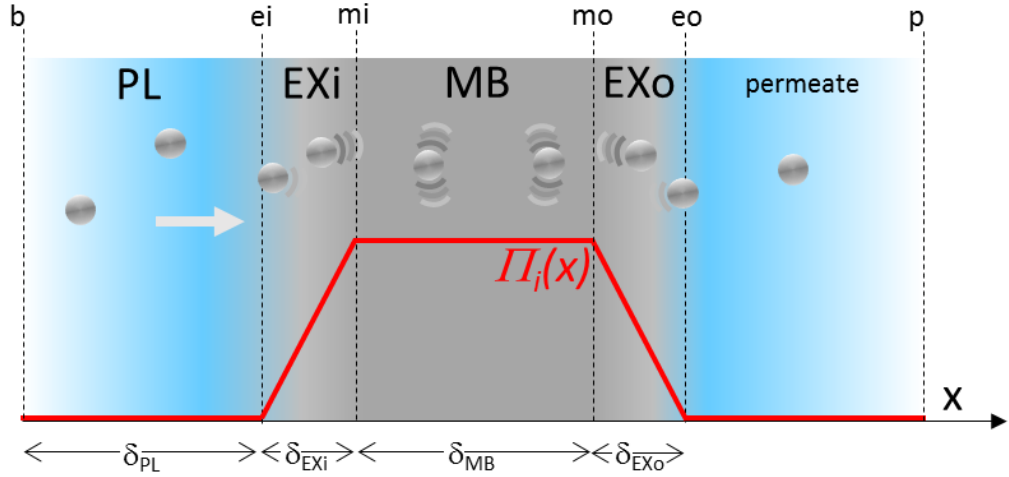

*Figure 2: The energy landscape representation of the interfacial pressure profile, $\Pi_i$, along the membrane thickness, x. The profile distinguishes the concentration polarization layer PL, the inlet exclusion layer EXi, the membrane core thickness, MB, and the outlet exclusion layer, EXo. The subscripts used for the value of the variables on these borders are b (bulk), ei (inlet exclusion), mi (membrane inlet), mo (membrane outlet), eo (exclusion outlet) and p (permeate).*

When applying the model (Eqs. 4-5) in these composite layers, the basis of physics are well translated. In the polarization layer, the model describes the equivalence between drag force and the gradient in osmotic pressure as already discussed by Wijmans et al. (1985) and Elimelech and Bhattacharjee (1998) for polarization concentration. As stated by Eq. 3, the pressure drop, $\nabla p$, is zero in the concentration polarization where there is no gradient in interfacial pressure, $\nabla \Pi_i = 0$. Furthermore, in the absence of drag force (i.e. at equilibrium, $u_m = u_c$ in Eq. 2), the equation set matches the description of the equilibrium between the static pressure and the osmotic pressure (Eq. 2 and 3 leads to $\nabla p = \nabla \Pi_{cc}$).

## 2.3 Modelling the transmission and the counter pressure through the membrane.

At a steady state and if considering a linear variation of interfacial pressure in the exclusion layers (Fig. 2), the application of the model enables the determination of analytical solutions. At a steady state, the mass flux, $j$, and the permeate flux, $J$, are constant through the membrane. The expression of the mass continuity (Eq. 4) in the composite layers determines the colloid concentration profile through the membrane (supplementary information S3). By assuming the continuity of the concentration through the composite layers, the ratio between the permeate concentration and the feed concentration (i.e. the colloid transmission, $Tr$) can be determined (supplementary information S4):

$$Tr = \frac{\phi_p}{\phi_b} = \frac{1}{1+Ex\left(\frac{(e^{-(Pe_{EXo}+Ex)}-1)e^{-Pe_{MB}-Pe_{PL}-Pe_{EXi}+Ex}}{Pe_{EXo}+Ex}+\frac{\left(1-e^{-Pe_{EXi}+Ex}\right)e^{-Pe_{PL}}}{Pe_{EXi}-Ex}\right)} \quad (6)$$

Where Péclet numbers, $Pe = \frac{J\delta}{D}$, are defined for each layers (with the subscript PL, Exi, MB and Exo) and, $Ex$, is the exclusion number defined in the previous section.

The permeate flow through the membrane, $J$, can be classically defined with a counter pressure, CP, describing the osmosis flow. When Eq. 5 is integrated, the counter pressure can be expressed as the integration of the interfacial force acting on all the particles in the exclusion layers $\int \phi d\Pi_i$ :

$$J = \frac{L_p}{\eta}(\Delta P - CP) = \frac{L_p}{\mu}\left(\Delta P - \int \phi d\Pi_i\right) \tag{7}$$

The counter pressure can then be analytically calculated from the energy map, $\Pi_i$, and the volume fraction profile, $\phi$, determined previously. In the end (the calculation details are given in the supplementary information S5), the counter pressure can be linked to osmotic pressures between the composite layers (Eq. 8) or to the operating conditions (Eq. 9) :

$$CP = \Pi_{ei} - \Pi_{eo} + \Pi_{mo} - \Pi_{mi} + \frac{Pe_{EXi}}{Pe_{EXi}-Ex}\left(\Pi_{mi} - \Pi_{ei} + \Pi_p Ex\right) + \frac{Pe_{EXo}}{Pe_{EXo}+Ex}\left(\Pi_{eo} - \Pi_{mo} - \Pi_p Ex\right) \tag{8}$$

$$CP = Tr\frac{\Pi_b Ex^2}{AB}\left(\frac{B}{A}(e^{-A}-1) + \frac{A}{B}(e^{-B}-1) + A + B - (e^{-A}-1)(e^{-B}-1)e^{-Pe_{MB}}\right) \tag{9}$$

where $A = Pe_{EXi} - Ex$ and $B = Pe_{EXo} + Ex$. This last equation is particularly interesting as it permits the counter pressure to be computed from ab-initio parameters. In contrast, the usual calculation (Eq. 1) requires the knowledge of the difference of the osmotic pressure through the membrane and the Staverman coefficient.

### 2.4 Model application and validation

To illustrate the ability of the model to describe membrane separation, calculations are now shown. Calculations are performed for conditions where the exclusion layers are small (1/10) compared to the others boundary layers: $Pe_{EXo} = Pe_{EXi} = 0.1 Pe_{MB} = 0.1 Pe_{PL} = 0.1 Pe$. The value of the partition coefficient is $e^{-Ex} = K = 0.1$. The transmission and the counter pressure determined with the analytical model (Eqs. 6 and 9) are provided in Fig. 3 and 4 as a function of the Péclet numbers, Pe (defined as $Pe = Pe_{MB} = Pe_{PL}$).

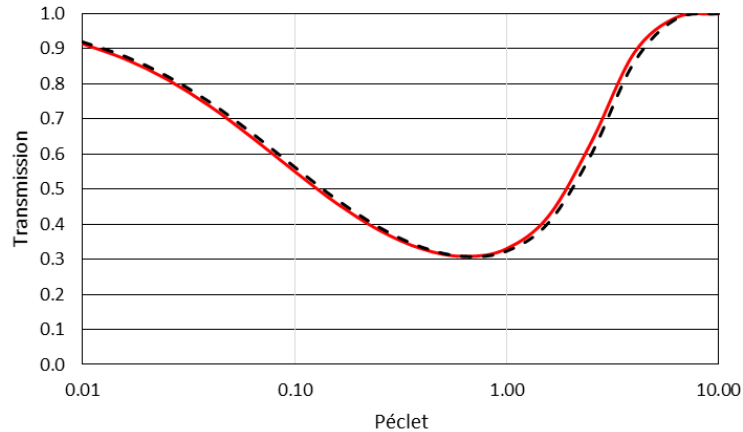

Figure 3 : Modelling of the transmission during a filtration for different Peclet number, $Pe = Pe_{MB} = Pe_{PL}$. The model (full red line) is compared to the classical transmission model of Opong and Zydney [36].

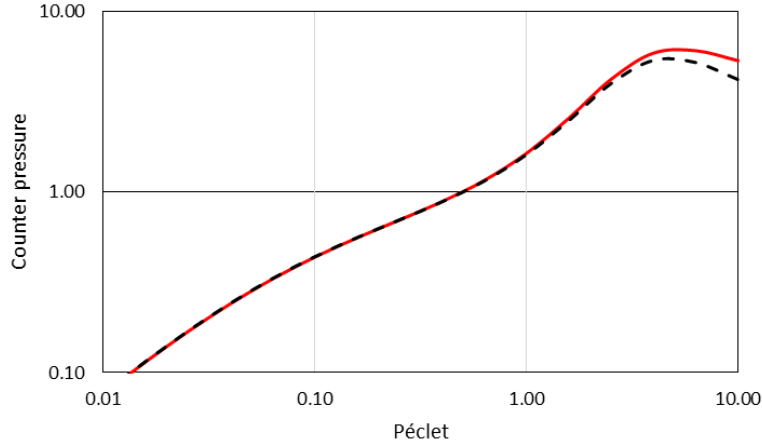

*Figure 4: Modelling of the counter pressure during a filtration for different Peclet number, Pe= $Pe_{MB}$ = $Pe_{PL}$. The model (full red line) is compared to the Kedem Katchalsky approach (dashed line) with a Staverman coefficient, $\sigma = (1 - K)$.*

It can be seen that the model for transmission (Eq. 6) provides results (Fig. 3) that are close to the classical relationship developed for membrane filtration [36,40] or for electrokinetic salt rejection [41]. The convergence of the model to the cases of an infinitesimally thin membrane is demonstrated in S6 of the supplementary information. Fig. 4 plots the values of the counter pressure as a function of the Pe numbers and compares the model to the classical approach, considering the osmotic pressure difference weighted with the Staverman coefficient. To operate this comparison, the Staverman coefficient [42] is defined as $\sigma = (1 - K)$ where K is the partition coefficient. Indeed, it can be demonstrated (detail in SI 7), after several simplification and model assumptions (for an infinitesimally thin membrane and for an osmotic pressure of colloids following the Van't Hoff law), that the counter pressure given by the model is converging toward:

$$CP = (1 - K)(\Pi_m - \Pi_b) \qquad (10)$$

This expression can also be seen as the contribution of $\Pi_{ei} - \Pi_{eo} + \Pi_{mo} - \Pi_{mi}$ in Eq. 8. Eq. 9 is therefore coherent with the frequent writing of the Staverman coefficient (Eq. 1) as a reflection coefficient accounting for the leakage of a membrane ($\sigma$ ranging from 0 for a completely non-retentive membrane to 1 for a membrane impermeable to the solute). The model therefore proposes a demonstration and analytical writing for the Staverman coefficient that is still subject to different interpretations of the Kedem and Katchaslky equations [43].

However, it must be noted, for the highest Péclet number (where conditions are highly irreversible and far from near-equilibrium assumptions), that there is a gap between the model and the Kedem Katchaslky Staverman approach, and the gap is amplified as the Péclet number increases (Fig. 4). These differences are due to the drag force that occurs in the exclusion layers. Indeed, the counter pressure results from the combination of an osmotic pressure difference and drag force (SI5):

$$J = \frac{L_p}{\mu}\left(\Delta p_f - (1-K)(\Pi_m - \Pi_b) - \int_{Exi+Exo}\frac{\phi}{V_p}F_{drag}dx\right) \qquad (11)$$

Eq. 11 can be seen as an extension of the Kedem and Katchaslky approach which accounts for the drag forces acting on arrested colloids. The drag force term is a frictional contribution that describes the resistance to the flow induced by the arrested (but still dispersed) colloids in the exclusion layers. Such a resistance may extend into a deposit flow resistance in the presence of a consolidated deposit (not taken into account in this paper). The drag force contribution is also represented by the last two terms in Eq. 8. This gap is becoming increasingly significant as the Peclet number increases (Fig. 2b) or when the thickness of the exclusion layers becomes thinner, as discussed in the following section.

# 3   Impact of energy landscape stiffness

In this section, the effect of the stiffness of the energy landscape on the membrane separation efficiencies is investigated. To do this, different energy landscape profiles are considered as sketched in Fig. 5.

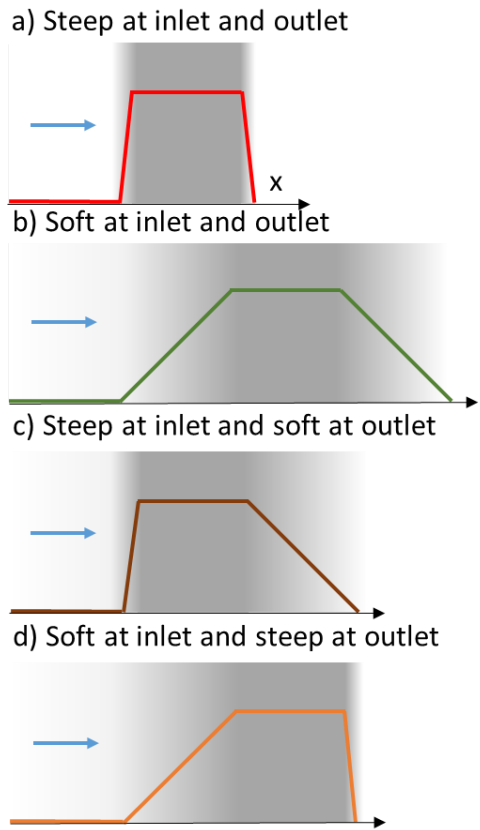

*Figure 5 : Energy landscapes considered for the analysis of the impact of the interaction stiffness. The steeper interactions occurs in thickness layers that are 1/10 of the membrane thickness. The softer interaction increases progressively in a thickness equal to the membrane thickness. For all cases, the thickness of the boundary layer is kept constant and equal to the membrane thickness.*

In section 3.1, the effect of the interaction stiffness on transmission is analyzed. Such effect can lead to an asymmetric transmission of the solute, discussed in section 3.2. The model is used to analyze experimental results of the asymmetric transmission of DNA plasmid through membranes. The stiffness of colloid-membrane interactions has also important consequences on the counter pressure and thus on the energetics cost of the separation that are discussed in section 3.3. Towards the end (section 3.4), these phenomena lead us to consider a paradigm shift to improve membrane efficiency: the accumulation at the membrane should be distributed along the exclusion layers of the membrane.

## 3.1 Impacts of colloid-membrane stiffness on colloid transmission

The model enables the investigation of the effect of the energy landscape profile on transmission (Eq. 6). To illustrate this effect, transmission and counter pressure were calculated for softer interaction ramps where the layers thickness for exclusion are in the same range as the membrane and the polarization layer thicknesses (Fig. 5b). In these conditions, the Péclet numbers are all equal, $Pe_{EXo} = Pe_{EXi} = Pe_{MB} = Pe_{PL}$, whereas for steep ramp (Fig. 5a and results in section 2.4), $Pe_{EXo} = Pe_{EXi} = 0.1 Pe_{MB} = 0.1 Pe_{PL} = 0.1 Pe$. Fig. 6 plots the transmission as a function of the Péclet number for these energy landscapes (Fig. 5). These results are compared to the relationship obtained when a partition coefficient depict the membrane exclusion (i.e. an infinitely thin exclusion layer –or infinitely steep interaction ramp-).

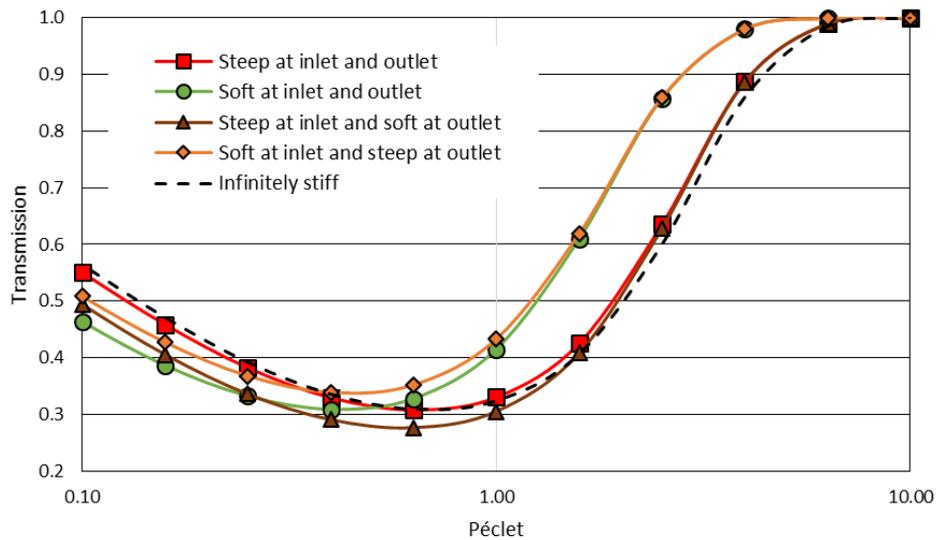

Figure 6 : The effect of the exclusion ramp stiffness on the transmission

For Péclet values greater than one, a softer interaction ramp stiffness leads to a shift in transmission towards smaller Péclet numbers. Thus, a similar transmission can be reached for smaller Péclet values (the colloids "climb" the membrane energy barrier more easily). In this Péclet range, the outlet ramp does not play an important role: the transmission is controlled by the stiffness of the interaction at the inlet. For Péclet values less than one (where diffusion plays a significant role in the transport), transmission is controlled by the total length of the membrane and the exclusion layers : an increase in the total length of the membrane(due to thicker exclusion layers) leads to a decrease in transmission of colloids.

## 3.2 Asymetric transmission through a membrane

The shift in transmission described by the model in the Fig. 6 has been observed in filtration experiments carried out in two different directions [44]:

- with the flow directed from the bulk to the skin layer of the membrane (the interaction occurs abruptly at the pore entrance) –referred to as forward filtration and corresponding to Fig. 5c-

- with the flow directed through the open pores in the substructure before the skin (the interaction with the pore wall take place progressively) –referred to as reverse filtration and corresponding to Fig. 5d-.

The experimental transmission of plasmid DNA obtained by Li et al. [44] in forward and reverse filtration are represented by symbols in Fig. 7. The results demonstrate that the transmission of the protein is facilitated when the filtration is performed in the reverse direction (Fig. 5d). The model (Eq. 6) has been used to interpret this data. The parameters of the model have been defined to have a good fit between modeling and experimental data as follows:

- the forward filtration (through the skin layer supported on the macroporous support Fig. 5c) is represented with a thin exclusion layer at the inlet and a thick exclusion layer at the outlet, $Pe_{EXi}$=0.1 Pe, $Pe_{EXo}$=5 Pe , $Pe_{PL}$=$Pe_M$=Pe
- the reverse filtration (through the macroporous layer first, Fig. 5d) is represented with a thick exclusion layer at the inlet and a thin exclusion layer at the outlet, $Pe_{EXi}$ =5 Pe, $Pe_{EXo}$ =0.1 Pe , $Pe_{PL}$=$Pe_M$ =Pe (thus corresponding to the reversal of the previous case).

For the fitting, the exclusion number is taken at 4.6 that corresponds, for infinitely thin membrane, to a partition coefficient of 0.01. The transmission experiments are well represented by the model (lines in Fig. 7). Furthermore, the fitting parameters are related to the physical conditions of the experiments: the Péclet number for the transport in the macroporous layer ($Pe_{EXo}$ in forward and $Pe_{EXi}$ in reverse direction) is 5 times higher than the one in the membrane. This proportionality coefficient is in the range of the ratio of the length of the macroporous layer ($\simeq$ 10 micrometers) and the length of the membrane skin layer ($\simeq$ 1 micrometer).

The enhanced transmission for the filtration in the reverse direction can thus be explained by the fact that the energy landscape is more progressive. The softer interaction ramp can be the consequence of the progressive spatial restrictions when filtering through the macroporous part of the membrane. The protein-membrane interaction are therefore progressively increasing. These interactions with walls can also lead to a change in the plasmid DNA conformation (as a plasmid extension) that can also in turn soften the energy landscape (as initially discussed in [44]).

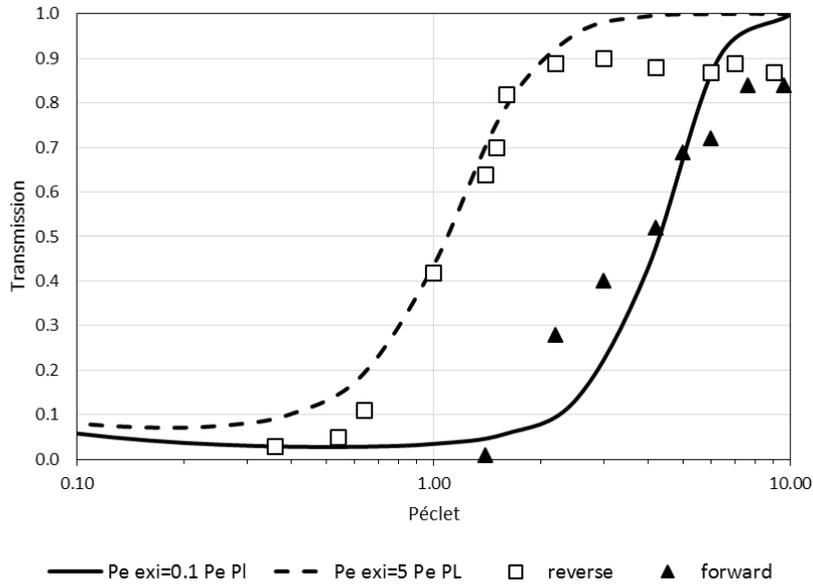

*Figure 7 : Protein transmission as a function of the Péclet number. The data is extracted from the paper of Li et al. 2015 for plasmid filtration through a membrane in the forward direction –from the skin layer to the macroporous layer- (full triangle) and in the reverse direction –from microporous to the skin layer- (open square). The lines represent the results obtained with the model. The full line is obtained for a thin exclusion layer at the inlet and a thick exclusion layer at the outlet ($Pe_{Exi}=0.1$ Pe, $Pe_{EXo}=5$ Pe, $Pe_{PL}=Pe_M=Pe$) thus representing the filtration through a skin layer supported on a macroporous support. The dashed line represents the results of the model for a thick exclusion layer at the inlet and a thin exclusion layer at the outlet ($Pe_{EXi}=5$ Pe, $Pe_{EXo}=0.1$ Pe, $Pe_{PL}=Pe_M=Pe$) corresponding to a case where the filtration is performed in the reverse direction.*

This non-symmetric membrane transmission displays a diode behavior by exhibiting varying resistance to the flow of protein according to the direction of use. Such behavior has been discussed [37] for nano-channels that have an asymmetry of surface charge (Fig. 1c). Better knowledge of such phenomena could help to develop new membrane design (with a porosity gradient for example) which could improve transfer efficiency and could also help to understand the asymmetric transmission observed with asymmetric biological cell membranes [45].

### 3.3 Softer colloid-membrane interaction ramp reduces the separation energy cost

Softer interactions also lead to a decrease in the counter pressure (Fig. 8) when the Péclet number is larger than one. For the same Péclet number, the counter pressure is lower when the interactions occur in a progressive (or softer) way.

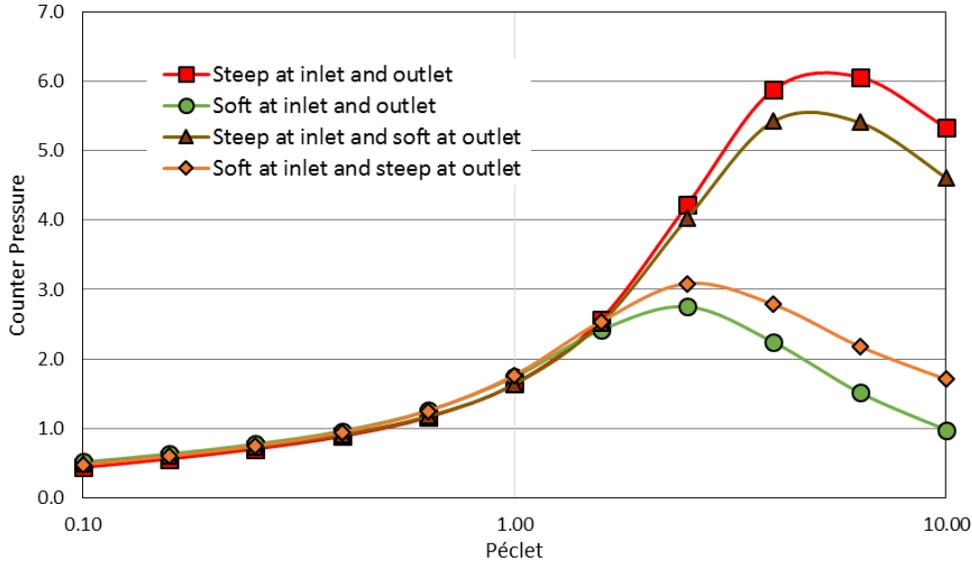

*Figure 8 : The effect of an exclusion ramp stiffness on the counter pressure*

However, it is difficult to fully interpret these results as the transmission is simultaneously varying with the Péclet number (as previously discussed with Fig.6 ). Thus, table 1 completes this data with values of counter pressure obtained for a same selectivity (i.e. for a same separation achievement). A same separation (a same transmission) can be achieved for 34% smaller Péclet number in the case of soft interaction ramps. As a result, the counter pressure is significantly reduced (-40 % when working with a 0.6 transmission or -55% for a 0.9 transmission).

*Table 1: Counter pressure, CP, values obtained for a same transmission, Tr, and for different conditions in the exclusion ramp stiffness. The counter pressure is compared to the difference of osmotic pressure, $\Delta\Pi$.*

|  | Tr | Pe | $V_p \Delta\Pi / kT$ | CP/$\Delta\Pi$ | CP (Pa) |
|---|---|---|---|---|---|
| Steep ramps | 0.6 | 2.36 | $4.24 \, 10^{-3}$ | 0.95 | 3.96 |
| Soft ramps | 0.6 | 1.56 (-34 %) | $1.90 \, 10^{-3}$ | 1.28 | 2.39 (-40 %) |
| Steep ramps | 0.9 | 4.11 | $6.09 \, 10^{-3}$ | 0.99 | 5.95 |
| Soft ramps | 0.9 | 2.79 (-34 %) | $1.63 \, 10^{-3}$ | 1.69 | 2.70 (-55 %) |

The counter pressure also represents the minimal energy in Joules required to produce 1 m³ of permeate (without accounting for additional solvent frictions inside the membrane). It is therefore interesting to plot the energy required to produce a permeate volume as a function of the process selectivity, *1-Tr*. Fig. 9 thus represents the consumption (that should be decreased) as a function of the selectivity (that should be increased). Moreover, to be economically sustainable, a separation process must achieve a productivity that is defined by the Péclet number (the values are displayed in text boxes in Fig. 9) that represents the flux intensity through the membrane.

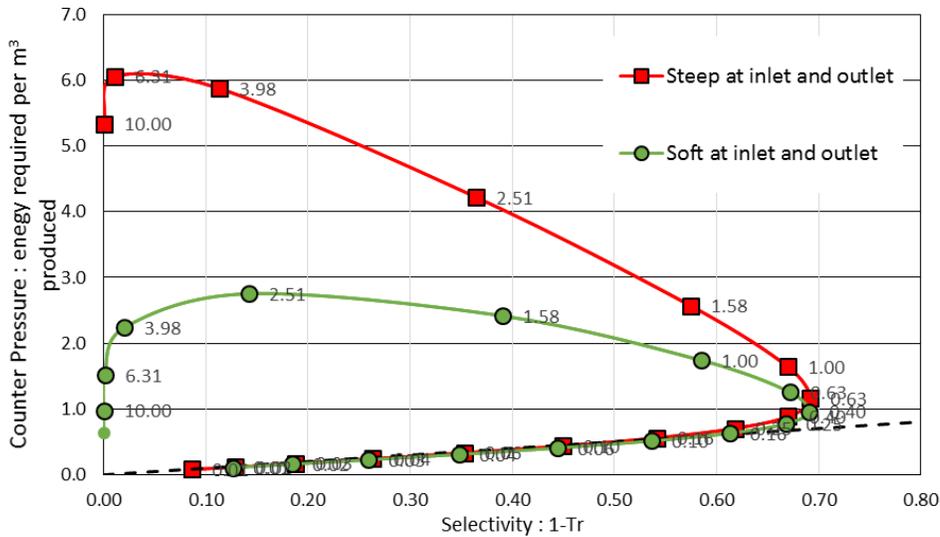

*Figure 9 : The benefit of a softer exclusion ramp on the separation efficiency (x axis) and energy costs required for the separation (y axis). Text boxes located close to the symbols give the Péclet value that indicates the productivity.*

For very low productivity (*Pe* value less than 1), the energy required for passing through the membrane is close to the bisector (dashed line) for which the counter pressure is equal to the concentration difference. However, when Péclet values are greater than 1 the energy required for the separation becomes significantly different according to the interactions conditions. This is mainly because of the polarization concentration and the drag forces in the exclusion layer (last term in Eq. 10) that dissipate more energy in these far from equilibrium conditions. For a same selectivity, the energy required for the transfer is less when the membrane interaction slope is softer mainly because of the reduction of these drag forces. Such condition corresponds to a more sustainable way to operate the separation. This could explain why hourglass shapes found in biological membrane transport channels can improve the transfer efficiency [5].

### 3.4  A new paradigm: the accumulation should be distributed

These results must help to show the membrane separation from a different point of view. The solute or the colloids transmissions through a membrane result from a potential barrier to overcome. The energy to overcome the barrier comes from the drag force (the permeate flux) that is transmitted on all the accumulated particles (the drag force must be multiplied by the number of accumulated layers or by the accumulated volume). Colloid transmission (controlling membrane selectivity) and colloid accumulation (controlling membrane productivity) are therefore fully linked. The colloid transmission is the consequence of the accumulation that allows providing enough energy to the colloids located close to the membrane to overcome the energy barrier and thus to pass through the membrane. The transmission of a colloid through the membrane results from the transmission of energy (due to drag force) through force chains in the network of accumulated colloids. But in usual membrane approaches everything is done to limit the accumulation. From this work, a change of paradigm can be considered. If considering that transmission needs some accumulation in order to take place, the accumulation is then a necessity to ensure transmission. But, if the accumulation occurs over a greater distance, a specific accumulated volume (controlling the transmission) can be reached without implying high concentrations (Fig. 10). The accumulation is distributed over a larger exclusion volume. As a consequence, for the same accumulated volume, the maximum concentration at the pore entrance is less and the counter pressure is decreased. Such configuration leads to a significant decrease in the energy cost for the separation. The main aim of the membrane user should be to distribute the accumulation, not to decrease the accumulation. A soft ramp in the colloid/membrane interaction is, for example, a way to distribute the accumulation and to perform filtration with a lower energy cost. Researchers [44] have experimentally demonstrated that such configuration could reduce fouling during protein filtration. The model can thus explain the gain due to pore hourglass shape, frequently existing in biological membranes.

Contrary to these conclusions, it must be noted that, classically, the design of membrane pores presents a sharp discontinuity in order to reject the accumulation outside the membrane. Indeed, the use of hourglass shape should be carefully considered and this new paradigm cannot be extrapolated to all the systems. This shape can represent an advantage, as previously discussed in this paper, only if the solute-wall or the colloid-wall interactions are highly repulsive to prevent the adhesion of the particles to the walls. For the opposite conditions, when this adhesion can occur, the hourglass shape can favors the clogging. It has been experimentally shown that a convergent flow at a microchannel bottleneck favor clogging by colloids [46]. Therefore, the hourglass shape should be combined with repulsive interactions (for example patchy interactions like with zwitterionic groups) avoiding the adhesion of the solute or the colloids to the wall. These theoretical findings could help to design ultrafiltration, nanofiltration or reverse osmosis membranes with a gradient of porosity (or free volume) in the skin layers and could help to elucidate the "strange" transport mechanism of fluids at the nanoscale [4].

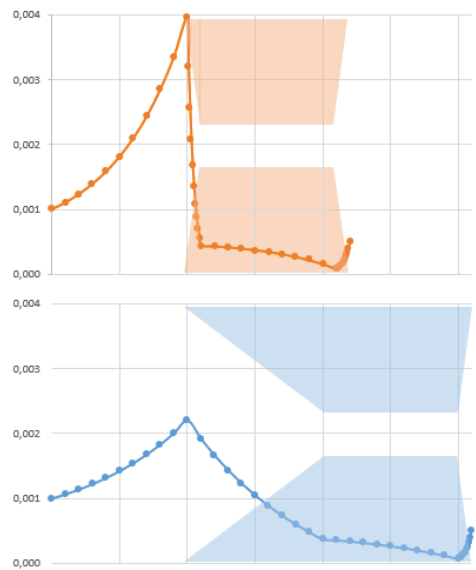

*Figure 10: Concentration profiles for a) steep interaction ramps (Fig. 5a) and b) soft interaction ramps (Fig. 5d). The Péclet values are 1.93 and 1.22 for a) and b) respectively. For both cases, the transmission is 0.5 and the exclusion number is 2.3. The soft interaction ramp leads to the distribution of the accumulated mass all along the interacting membrane. This hourglass shape for the interacting membrane leads a reduction in the maximum volume fraction, thus reducing the counter pressure (the counter pressure are 3.19 and 2.08 Pa for a) and b) respectively). The efficiency of the transfer is therefore higher in the case of soft interaction profiles.*

# 4  Conclusions

A model, based on a two-fluid approach, has been solved in an energy landscape to account for the colloid/membrane interactions. Colloid transmission and the counter pressure can be analytically determined from ab-initio operating parameters. The model provides the theoretical origin of the empirical writing of the Darcy law with the effective pressure given by the gap to the equilibrium, $\Delta P - \sigma \Delta \Pi$. The Staverman coefficient, $\sigma$, is also theoretically linked to the partition coefficient of the membrane and a gap to the Staverman approach due to drag forces is evidenced in far-from equilibrium conditions. The ability of the model to describe the transport in an energy landscape is illustrated by the investigation of the effect of a ramp of interaction. These results show that the presence of progressive interactions increases the efficiency of the transport and decreases the counter pressure. These results introduces a new paradigm: the accumulation should be distributed along the membrane thickness, not decreased, to increase the separation efficiency. This work can explain the results that show that filtration in a reverse direction (through the macroporous support) can be more efficient than in a forward direction (through the skin). It gives a possible explanation for the effect of the hourglass shape in membrane pore channels and for the membrane diode behavior. It opens up interesting potentialities for the optimization of the membrane porosity design and for the study of dynamic transport in pores with a nanoscale interaction mosaic, as for biological membranes.

# 5  Acknowledgements

The author thanks Yannick Hallez, Fabien Chauvet, Leo Garcia, Martine Meireles and Pierre Aimar for the fruitful discussions on the model.